# Design and Implementation of Air Selection based Augmented Reality Serious Game for Learning Capability Analysis

Harini. M, Harini. T, Roxanna Samuel

*Abstract: Rising advancements and ICT have changed the way of life of society, every single logical zone are exploiting innovation to get a genuine improvement. Specialists understand the advantages of utilizing genuine games as a dependable device in psychoanalyst. Hence, the exploration looks at important issues in regards to Dyspraxia issue in youngsters and presents a similar report in the treatments strategies by utilizing a non autonomous riddle and by utilizing the game, a Serious Game created in the intension of helping kids suffering from Dyspraxia to enhance their engine aptitudes and deftness through innovation. The investigation of information results indicated that exist a critical distinction among the two strategies, demonstrating that youngsters spending time with Serious Game got little schedule in the movement running and furthermore enhanced execution.*

## I. INTRODUCTION

As per WHO Worldwide Classification of Diseases, there is a few meanings of learning handicaps, and some of those people need the individual to have an IQ under 70 [1]. These handicaps can likewise meddle with more elevated level aptitudes, for example, association, time arranging, conceptual thinking, long or momentary memory and consideration. It is fundamental to know that learning handicaps can influence a person's life past scholastics and can affect connections [2].

It is remarkable that each individual gets a ton of hereditary report carried retrieved from people & when individual's are growing up, their condition and experiences also shape their properties & improvement. There is larger number of adolescents than adults with learning handicaps. Studying inadequacies are ordinary. These adolescents need extra help at institution to acquire the most clear opportunity to learn scholarly limits [1].

Considering inadequacy are result of differentiation on the manner that an individual's cerebrum is "wired." Individuals, youths with studying insufficiencies is as shrewd lke their friends. In any case, they may encounter issues scrutinizing, spelling, making, thinking, surveying and sifting through information at whatever point left to comprehend things without any other person's info [2]. With the best possible assistance and intercession, kids who has studying insufficiencies win at institution.

Taking in failures are really factor beginning with one child then onto the following. One individual could fight with scrutinizing and logography, while the other loves books anyway can't get math. Then again potentially another child may encounter issues understanding what others are expressing or bestowing for all to hear. The issues are inside and out various, anyway they are generally studying ineptitudes [4, 5]. A bit of the learning failures are APD and ADHD.

In this particular logical examination, the Dyspraxia insufficiency were inspected due to the risks it conveys through thought inadequacy into posterity with specialist age. The engine trouble hints problems with progression and arranging.

Signs show that an adolescent may have a motor systemization inadequacy consolidate issues with real limits that require dexterity [1, 6-8]. Adolescents may show poor evening out; will appear to be ungraceful; will intermittently falter. Moreover, these show issue during motor orchestrating, shows deficiency in the ability to sift through self and resources, exhibits possible affectability to contact, might break things or pick dolls which don't desire gifted control. They experience issues with fine motor assignments, for instance, concealing amidst the lines, fragmenting absolutely, collecting mazes, or sticking splendidly, disturbed by scratches, cruel, stiff or significant clothes.

Nowadays, new ICT and creating progressions open on the planet, a couple of pros have proposed interventions through gamified sorted out activities [9]. At the present time, has been a basic progression of medicines for different failures reliant on informational entertainments called as SG.

Serious Game explanation had moved after some extent; regardless, the most dependable variation was established by Clark C. Abt, in the 1970 document with title "Real Games." He says: "We are stressed over veritable entertainments as in these games have an express and intentionally thought about explanation and aren't used to be rejoiced fundamentally for beguilement." Serious Games are normally seen as promising training and studying gadgets for the Twenty First century [13, 14].

The principle rule disputes is that games are speaking to nowadays understudies [14]. Besides, the basic issues and challenges are making serious games together with development improved studying draws near, for instance, Augmented Reality (AR). At this moment, headways are an amazing assurance considering the way that they spike kids with new challenges; giving a brisk information, that is custom fitted into the particular involvement and people's
needs.

These ARSG must have the alternative to get their thought & improves correspondence technique with experimentation and multiplication constrained by strong-arm advancements.

A basic need exists to design and propose inventive drugs using development to help kids with Dyspraxia that acclimates to the necessities through the development subject to the typical UI. It could offer boundless possible results are progressively customary joint effort, that can see body and hand movements, making this an ideal gadget to outfit imaginativeness into medicinal activities got together with ARSG.

The basic objective of this assessment is to grandstand the outcomes got after an arranging technique with two parties of youngsters utilizing an ARSG model. It is orchestrated by appropriate intervention rehearses set up by specialists in the play territory of Dyspraxia to improve deftness limits, info, nature, and essential thinking. This model mishandled the AR and trademark interface utilizing Three Dimensional virtual conditions and Kinect, this consistent assessment assisted with studying the adequacy of ARSG.

Right off the bat, general data identified with dyspraxia, genuine games and enlarged the truth was presented right now. Area 2 exhibits the ARSG configuration process. Segment 3 portrays the exploration technique.

## II.    SERIOUS GAME DESIGN PROCESS

For this assessment , 4 stages were put into consideration(examination, plan, improvement, and evaluation) on the item life process [19]. The assessment arrange incorporates examination of the necessities, thinking about the circumstances, scholastic perspectives, learning substance and vivacious. This method chooses a ton of stages whose basic objective is to perceive the different parts of the formation of the SG. At this moment criteria were developing [22]:

- Objective for the game
- Entertainment additional items.
- Prizes.
- Necessary kid focused substance
- Entropy to give shock

Relative studying exercises, 5 standards were picked:
- Establishment of the studying material
- Shaping input on studying
- Reserve language (even as pictures)
- Eubstance of studying introduction.
- Interlinkage.

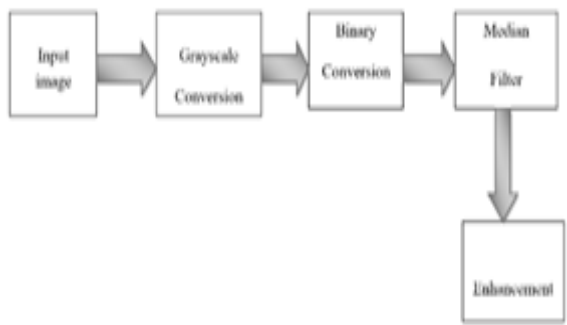

**Figure 1. Image Preprocessing Block Diagram.**

In the arrangement arrange, modernized resources imperative to development of the serious games must be made, including Two Dimensional and Three Dimensional plots, sorted out articles, sounds that reflect the examination subtleties. Moreover, portrayed the interrelatedness betwixt educational substance & planning. This stage underscores the family relationship of the instructive objectives and the difficulties of the game, which are grown emphatically. It depicts the principles and instruments of play game. Kinect connects with gamers to control and group up, utilizing a trademark UI.

• Arduino uno:
For transmitting and getting signals.
• Three Ultrasound sensors:
Air development of hand determination is caught through the sensors and which will be brought for examination.
• Software Used:
MATLAB code to structure a GUI indicating the animation character pictures in a 3*3 cluster position.

For the visual depiction, AI permitted the symbolism advancement, AAE programming were utilized to actualize/model activities &  thereby advance yield introduction. At long last, Adobe Premiere Pro [27] permitted altering proficient recordings.

For the production of AR app, the SDK Vuforia  was selected, This  empowers the different gadgets combine the game's interfaces. Kinect for Win DK  was likewise done to consolidate every single game component using a characteristic UI. Each one of those devices, under an incorporated advancement condition, utilizing the high hat programming rehearses.

The serious games assessment stage is supplemented by 2 jobs: the end client & the master, which unites various viewpoints. Right now, approval, input, & testing method were checked.

### A.    Game Characters

SERIOUS GAME have 9 chars; every one is spoken to by a symbol of recognizable proof and finishes activities. A situation
contains scenes and mixed media components, which are a piece of restorative difficulties. Mention that characters are notable characters which are attractive for children and motivates to play
the game. (Figure 1).

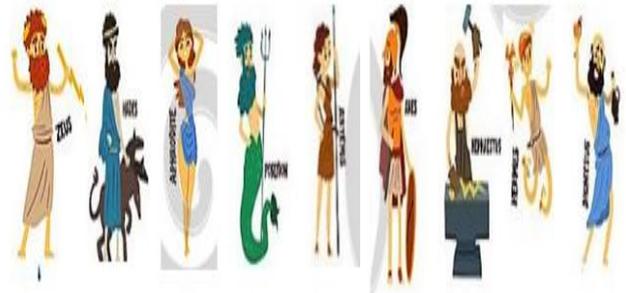

**Fig 2. Serious Game Characters.**

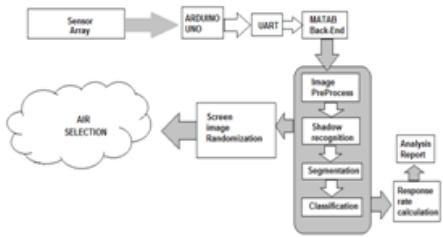

Fig 3. Functional Architecture Diagram

### B. How to play SERIOUS GAME

1. The advisor discloses the system to begin the game.
2. Each youngster needs to login into the game through a symbol determination. There are six symbols, which speak to kids from Ecuadorian locales. Every kid must pick one and afterward compose his/her name utilizing a characteristic interface.
3. A cinematic is seemed explain the chars present in SERIOUS GAME.
4. After that, a menu of situations is shown, which substance diverse sort of treatments. There are three levels (novices, intermedium and propelled), the trouble level compares to the aptitudes and abilities of the player.
5. The treatment appeared in Figure 2 tends to Shapes situation. The objective of this scene is to compose the chars on the correct face with right form on the left side. Inside the game, a database spares registers of time, achievement, and disappointments happened during the game mode.
6. Results will be displayed.
7. Finally, input sessions were arranged.

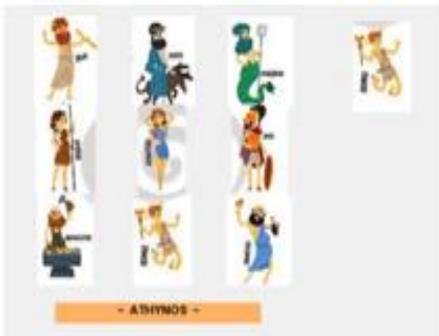

**Fig 3. Screen capture of Shapes situation in SERIOUS GAME.**

### III. STRATEGY RESEARCH

#### A. Participants

Study members were 45 youngsters (20 young men and 20 young ladies), (M= half; F=50%) isolated into two gatherings in haphazardly and freely way to maintain a strategic distance from conceivable inclination in the example (20 kids in each gathering). The age normal is 8.4 years old (SD=0.84).

To do an examination, the main gathering was called Control Group (CG), under a lone customary educating learning technique and then again, the subsequent gathering, called Experimentation Group (EG), utilizing SERIOUS GAME. Focuses' executive got the parent's composed consent for kids' interest right now. For now experimentation, two specific analyst/advisors had been answerable for leading and applying intercession sessions.

Every youngster was welcome to go to 4 sessions, in each gathering a non autonomous riddle treatment & SG model were utilized. Furthermore, an arbitrary request of youngsters' cooperation was built up in every session, which kept going roughly 20 minutes. The manual riddle treatment was an academic game for sensorimotor training, and it comprised on to coordinating letters and words with figures as per the advisor guidelines.

Time of exercises execution, and execution were enrolled by every understudy, considering the evaluation size of the Education Ministry of Ecuador (1-10), where 1 methods the most reduced evaluation, and 10 is the best one. See Table I.

**Table I. Grades scale of the Education Ministry of Ecuador**

| Grades | Meaning |
|---|---|
| 10 | Surpasses the learning |
| 9 | Ace the learnings |
| 7-8 | Accomplishes the necessary learning |
| 5-6 | Is near accomplishing the learning |
| ≤ 4 | Doesn't accomplish the necessary |

#### B. Procedure

At each session, a chance to light up the consigned practices by youths were noted. With these information, the time run of the mill of every single get-together for every kid was settled. After that a quantifiable evaluation was made. Utilizing the SW test, it was discovered that information dispersing isn't average. That was the inspiration driving why the SW procedure was used since there was an immense differentiation among the time apportionments, securing a p-esteem = 6.748e-0859<0.0545 (Table II). These assessment between time delayed consequences of the two medications is shown in Figure 4 and Figure 5.

**Table II. Time measurement outline(Min**

| Method | Min. | 1st | Median | Mean | 3rd | Max. |
|---|---|---|---|---|---|---|
| Non Autonomous | 4.54 | 4.698 | 4.855 | 4.798 | 3.923 | 4.998 |
| SERIOUS GAME | 3.024 | 3.345 | 3.645 | 3.595 | 2.955 | 4.018 |

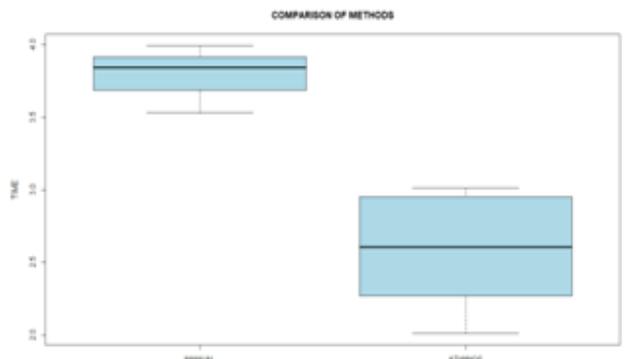

**Fig 4. Comparison results of Distance and Time**

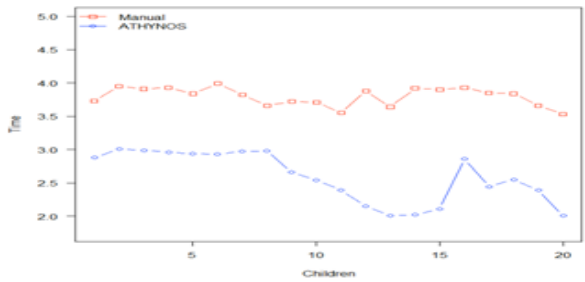

**Fig 5. Scoping examination of Time.**

An engraving given for individual understudy during every meeting as an assessment of young people, contemplating the correct answers and thwarted expectations. Beginning there ahead, the assessments is enrolled, and with the information, midpoints were settled for each understudy, and are brought right now scholarly execution.

Along these lines, in a fundamental way, utilizing the SW test, it is conceivable to set up that spread of the information isn't ordinary. That was the inspiration driving why the Wilcoxon system was used since there was a basic differentiation among the assessments dispersal, getting a p-esteem =0.0001078 < 0.545 (Table III).

**Table III. Understudy Performance measurement outline**

| Method | Min. | 1st Quat. | Median | Mean | 3rd Quat. | Max. |
|---|---|---|---|---|---|---|
| Manual | 7.250 | 7.450 | 8.200 | 8.023 | 7.500 | 9.450 |
| SERIOUS GAME | 8.350 | 7.250 | 9.225 | 8.012 | 8.250 | 9.450 |

The correlation between execution aftereffects of the two treatments is introduced in Figure 5 and Figure 6. The measurable examination was made utilizing R programming in the two cases with time and scholastic execution.

C. Results

The boxplot and the distinct examination of information affirm t the interaction time of kids' exercises is longer when they work with non autonomous treatment exercises. In the interim, when kids utilized SERIOUS GAME game, there was a noteworthy abatement in the time utilized by youngsters at the treatments. Accordingly, there was an enhancement in their engine range and hand-eye coordination dependent on boxplot of execution. Additionally, it was seen that the fluctuation of the occasions got by kids was homogeneous in the two ocurrences, which shows that all kids have comparative volumes in two approaches.

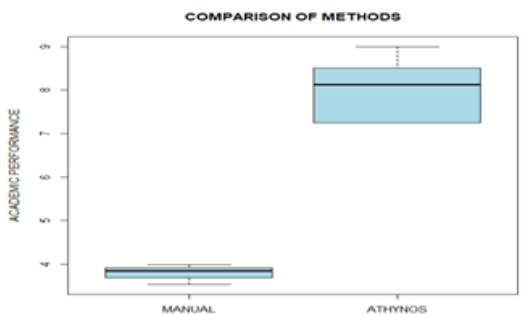

**Fig 6. Examination of interaction Allocations**

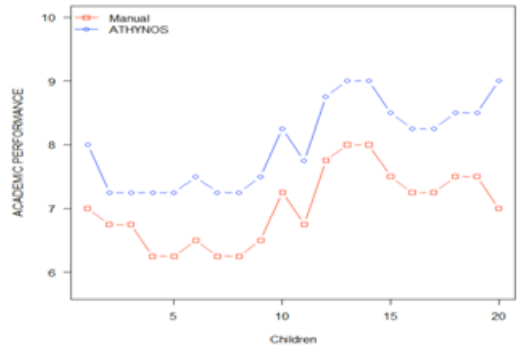

**Fig 7. Near investigation of System Performance between CG and EG.**

IV. CONCLUSIONS

The essential objective of this investigation was to improve youngsters engine aptitudes through SERIOUS GAME game. It causes youngsters to be progressively occupied with physical preparing and improving their substantial sensation knowledge considering that kids are advanced locals. They got the opportunity to secure a specific example of reasoning and be increasingly happy with fine engine abilities.

The proposed look into technique with respect to the occasions and execution given by SERIOUS GAME model introduces measurably critical enhancements at 95% certainty concerning manual strategy. The kids improved their finding out about dexterity, reciprocal joining, and sequencing. Likewise, SERIOUS GAME give criticism on gamers accomplishments and let realize that exercises are doing effectively. It is significant since it propels kids to address the difficulties set in the game.

A few investigations have proposed mediations of treatments for various handicaps dependent on rising advances (AR) utilizing regular UI.

For next level, creators contemplate improving the game incorporating new ranges & exercises. Also, it is essential to test SERIOUS GAME with more offspring including various trial bunches so as to approve the fundamental outcomes acquired right now.

## AUTHORS PROFILE

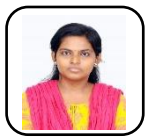

**Harini M**, is currently pursuing her Bachelor of Engineering degree in Computer Science and Engineering at Rajalakshmi Engineering College, Anna University. She has Completed 5 Mini Projects on HTML, CSS and Javascript with database connectivity. Her areas of interest include Iot, Cloud Computing, Big Data, Web Development and BPM. She has published her research paper in the domain of Block Chain Technology. She is also a Pega Certified Senior System Architect(PCSSA).

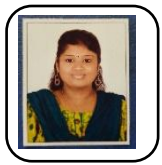

**Harini T**, is currently pursuing Bachelor of Engineering degree in Computer Science and Engineering. She has completed three mini projects on HTML, CSS and JavaScript with database connectivity, JAVA and mobile application development. Her areas of interest include HTML, CSS, JavaScript and Java. She has been certified in Cloud computing and Bigdata

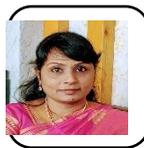

**Mrs. Roxanna Samuel** has completed her Bachelors in Computer Science and Engineering at Madras University in 1999 and Masters in Computer Science and Engineering at Deemed University in 2004. She is presently working as Assistant Professor (SG) in the Department of Computer Science and Engineering at Rajalakshmi Engineering College, Chennai. She has published many research papers in various journals and conferences covering several domains like, Big data, IoT, Data Mining. She has more than 12 years of teaching experience in various Engineering Colleges. Her area of interest includes Networks, IoT, Digital Technology, Data Analytics, Cryptography and Network Security. She has guided and is guiding students in both UG and PG Level and received the best faculty award in Computer Science and Engineering.